\documentclass{emulateapj}

\newcommand{\pga}{PG 0112$+$104}

\slugcomment{Submitted to The Astrophysical Journal}
\shortauthors{Dufour et al.}
\shorttitle{Multiwavelength Observations of \pga}

\begin{document}

\title{Multiwavelength Observations of the Hot DB Star \pga\altaffilmark{1}}

\author{
P.~Dufour\altaffilmark{2}, S.~Desharnais\altaffilmark{2}, 
F.~Wesemael\altaffilmark{2}, P.~Chayer\altaffilmark{3}, 
T.~Lanz\altaffilmark{4}, P.~Bergeron\altaffilmark{2},
G.~Fontaine\altaffilmark{2}, A.~Beauchamp\altaffilmark{2,5},
R.A.~Saffer\altaffilmark{6}, J.W.~Kruk\altaffilmark{7},
M.-M.~Limoges\altaffilmark{2}
}

\altaffiltext{1}{Based on observations with the FUSE satellite, which
is operated by the Johns Hopkins University under NASA contract NAS
5-32985; with the NASA/ESA {\it Hubble Space Telescope}, obtained at
the Space Telescope Science Institute, which is operated by the
Association of Universities for Research in Astronomy, Inc., under the
NASA contract NAS 5-26555; and with the Kitt Peak National Observatory,
National Optical Astronomy Observatories, operated by the Association
of Universities for Research in Astronomy, Inc., under cooperative
agreement with the National Science Foundation.}
\altaffiltext{2}{D\'epartement de Physique, Universit\'e de Montr\'eal,
C.P. 6128, Succ. Centre-Ville, Montr\'eal, QC H3C 3J7, Canada;
dufourpa, stephanie, wesemael, bergeron, fontaine, limoges@astro.umontreal.ca}
\altaffiltext{3}{Space Telescope Science Institute, 3700 San Martin Drive,
Baltimore, MD 21218, USA; chayer@stsci.edu}
\altaffiltext{4}{Department of Astronomy, University of Maryland,
College Park, MD 20742, USA; lanz@astro.umd.edu}
\altaffiltext{5}{Forensic Technologies Wai Inc., 5757 Boulevard
Cavendish, Montr\'eal, QC H4W 2W8, Canada;
alain.beauchamp@fti-ibis.com}
\altaffiltext{6}{DataTime Consulting, 109 Forrest Avenue, Narberth, PA
19072, USA; rex.saffer@comcast.net}
\altaffiltext{7}{Bloomberg Center for Physics and Astronomy, The Johns
Hopkins University, Baltimore, MD 21218, USA; kruk@pha.jhu.edu}

\begin{abstract}

We present a comprehensive multiwavelength analysis of the hot DB white
dwarf \pga. Our analysis relies on newly-acquired FUSE observations,
on medium-resolution FOS and GHRS data, on archival high-resolution GHRS
observations, on optical spectrophotometry both in the blue and around
H$\alpha$, as well as on time-resolved photometry. From the optical
data, we derive a self-consistent effective temperature of $31,300\pm
500$ K, a surface gravity of log $g = 7.8\pm 0.1$ ($M=0.52\,M_\odot$),
and a hydrogen abundance of ${\rm log}\,N({\rm H})/N({\rm He}) < -4.0$.
The FUSE spectra reveal the presence of \ion{C}{2} and \ion{C}{3} lines
that complement the previous detection of \ion{C}{2} transitions with
the GHRS. The improved carbon abundance in this hot object is ${\rm
log}\,N({\rm C})/N({\rm He}) = -6.15\pm 0.23$. No photospheric features
associated with other heavy elements are detected. We reconsider the role
of \pga\ in the definition of the blue edge of the V777 Her instability
strip in light of our high-speed photometry, and contrast our results
with those of previous observations carried out at the McDonald Observatory.

\end{abstract}


\section{Astrophysical context}

The DB stars constitute the subgroup of helium-atmosphere white dwarfs
whose optical spectrum is dominated by the transitions of neutral
helium. Their ef\/fective temperatures extend from roughly 13,000 K
upward to $\sim 40,000$. The lower boundary is imposed by the
visibility of the He I lines, that become very weak and disappear near
that ef\/fective temperature, while the upper boundary is the
effective temperature near which the smooth merging occurs between the
DB sequence and the sequence of the hotter DO stars, whose spectrum is
characterized by lines of ionized helium. Some filling-in of the
region between 30,000 K and 40,000 K, previously thought to be devoid
of helium-atmosphere white dwarfs \citep{wgl85} and known as the ``DB
gap''\citep{lfw87}, has been accomplished with the recent observation
by \citet{eisensteinetal06} of several fainter hot DB stars in the
Sloan Digital Sky Survey (SDSS).

While the idea of a true gap in the cooling sequence of the
helium-atmosphere white dwarfs has now been abandoned,
\citet{eisensteinetal06} argue nevertheless for the presence of a
residual imbalance in the relative numbers of DA and DB stars, in the
sense that the DA/DB number ratio is $\sim 2.5$ times larger at 30,000
K than it is at 20,000 K. This result is interpreted in terms of a
transformation of $\sim 10$ \% of the DA stars observed at 30,000 K
into DB stars by the time these objects cool down to 20,000 K. If this
were the case, the physical mechanism invoked by \citet{lfw87} and
\citet{fw87} within their global spectral evolution scheme for white
dwarfs might still be relevant. It would require that a fraction of DA
white dwarfs be characterized by a hydrogen envelope that is thin
enough (of the order of $M_H < 10^{-14.9}\,M_\star$) for mixing to
start between that envelope and the active, underlying helium
convection zone. The photospheric composition would thereby revert to
a helium-dominated one as the star proceeds along its cooling track.

Before the numerous SDSS discoveries, the northern DB star \pga\ had
long held the status of the hottest DB star. Today, it merely appears
cooler than most of the SDSS stars. Nevertheless, its discovery in the
PG survey at a reasonable brightness ($V\sim 15.4$) makes it
accessible to a variety of instruments, a substantial bonus for a hot
DB star. As such, it represent the best studied DB white dwarf above
30,000 K. Since it is not exactly understood how DB stars above 30,000
K fit into white dwarf evolution, a comprehensive analysis of \pga\ is
crucial for an understanding of the still uncertain nature of the
spectral evolution mechanism. \pga\ is also the best object near the
uncertain boundaries of the V777 Her variable separating hot DB star
from DB pulsators so an accurate determination of its atmospheric
parameters is important for an unambiguous empirical location of the
blue edge of the DB instability strip. 

Prompted by these considerations and by FUSE observations secured by
one of us (TL), we endeavor here to reexamine \pga. With the recent
demise of FUSE, this object is the last classical DB star for which
FUSE data are available and unpublished. These data are combined here
with additional, and largely unpublished, information available on
this object in order to constrain ef\/fectively its atmospheric
properties and to try to shed light on its nature. In addition, the
status of \pga\ as a non-variable star has recently been challenged
\citep{shipmanetal02,jlp03} so we take advantage of this study to
reconsider the issue of the variability of this critical object.

We present the observational material on which this investigation is
based in \S\ 2, and discuss some issues related to the modeling of DB
stars in \S\ 3 and 4. Our reanalysis follows in \S\ 5, while our
conclusions are presented in \S\ 6.

\section{Observational material}
 
\subsection{FUSE data}
 
The FUSE detectors span the ultraviolet region between 905 and 1187
\AA, a region where numerous transitions associated with heavy
elements are located. The spectroscopic observations of \pga\ were
secured in TTAG mode through the high-throughput LWRS aperture, which
provides high-resolution ultraviolet spectra with
$\lambda/\Delta\lambda \sim$ 15,000. One set of exposures, 22.9 ks
long, was secured on 2004 January 2 within GO program C026 (P.I.: TL),
while a second, shorter series of exposures (4.25 ks) was secured by
JWK on 2004 December 19 within the FUSE photometric calibration
program M102. Table 1 summarizes these spectroscopic observations,
together with those secured for this project with other
instruments. Data were reduced using the pipeline software CalFUSE
(version 3.2.1). A detailed discussion of CalFUSE 3 is presented by
\citet{dixonetal07}. The individual exposures were coadded for
segments A and B for the four channels (SiC1, SiC2, LiF1 and LiF2)
individually. To carry out this procedure, we use the publicly
available program FUSE\underbar{ }register written by
\citet{lindler01} together with scalar weights based on the exposure
times. The eight weighted segments were then coadded with the use of a
linear interpolation between the spectra in order to produce a final
spectrum covering the whole FUSE spectral range. Figure 1 shows the
whole FUSE spectrum of \pga\ that was obtained by merging the SiC1B,
LiF1A, SiC2B, LiF2A, and LiF1B segments. For the spectral analysis, we
combine the segments that provide the best resolution after accounting
for the small wavelength shifts between them. In all cases, the noise
level varies throughout the spectrum. This is due to the significant
change in the ef\/fective area of the channels with wavelength, which
results in variations in the signal-to-noise ratio. In addition, the
number of segments covering a given spectral region is a function of
wavelength.

\begin{deluxetable*}{lrccccc}
\tablewidth{0pc}
\tablecaption{Summary of FUV, UV, and Optical Spectroscopic Observations.
\label{tab:log_spec_obs}}

\tablehead{
\colhead{Program ID} &
\colhead{Date} &
\colhead{Grating} &
\colhead{$R\equiv\lambda/\Delta\lambda$} &
\colhead{Wavelength (\AA)} &
\colhead{Exp. Time (s)} &
\colhead{Instrument} 
}
\startdata
C0260301 & Jan 02 2004 & \nodata &18,000& 905--1187 & 22899 & {\it FUSE} \\
M1020101 & Dec 19 2004 & \nodata &18,000& 905--1187 & 4250 & {\it FUSE} \\
Z3EU0106P & Nov 15 1996 & G140L & 2000 & 1136--1422 & 653 & $HST$/GHRS \\
Z3G70207T & Dec 02 1996 & G160M & 15,000 & 1196--1234 & 5658 & $HST$/GHRS \\
Z3G60205M & Nov 14 1996 & G160M & 17,000 & 1316--1353 & 3699 &$HST$/GHRS \\
Z3EU0107P & Nov 15 1996 & G140L & 2300 & 1389--1676 & 1197 & $HST$/GHRS \\
Y3EU0103P & Nov 15 1996 & G190H & 1300  & 1571--2311 & 480 & $HST$/FOS \\
Y3EU0104P & Nov 15 1996 & G270H & 1300  & 2222--3277 & 160 & $HST$/FOS \\
\nodata   & Jul 19 1991 & 600(1) & 700   & 3750--5100 & 2100 & 2.3~m/B \& C \\
\nodata   & Nov 23 2009 & 600(1) & 600   & 3200--5300 & 1200 & 2.3~m/B \& C \\
\nodata   & Sep 29 1999 & KPC 18C & 2200   & 5600--7400 & 2400 & 4~m/R-C \\
\nodata   & Feb 24 2000 & KPC 18C & 2200   & 5600--7400 & 1800 & 4~m/R-C \\ 
\enddata
\end{deluxetable*}

\begin{figure}[!ht]
\plotone{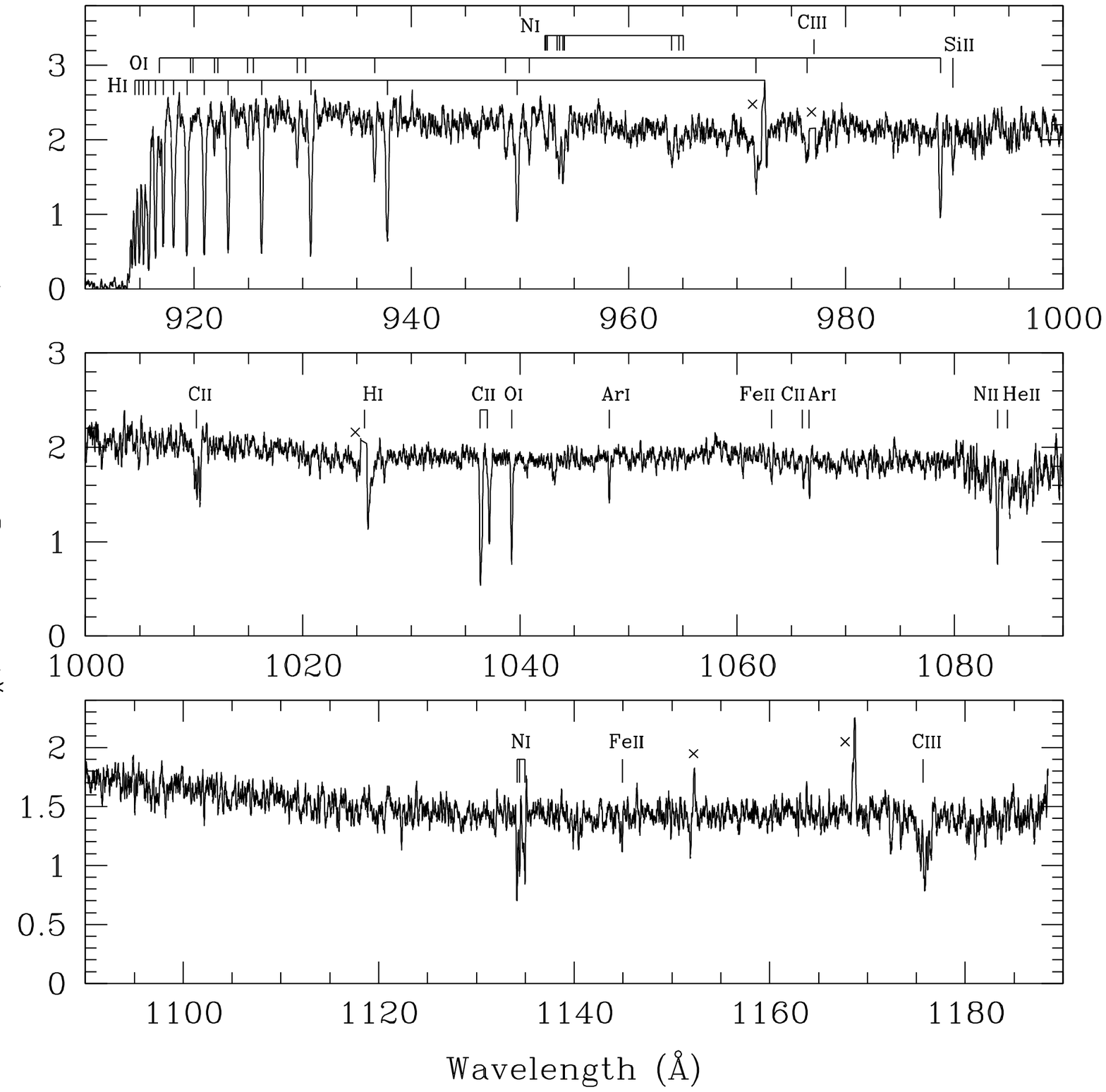}
\caption{FUSE spectrum of \pga. The spectrum is
obtained by merging the SiC1B, LiF1A, SiC2B, LiF2A, and LiF1B
segments.  The interstellar \ion{H}{1}, \ion{N}{1}, \ion{N}{2},
\ion{O}{1}, \ion{Si}{2}, \ion{Ar}{1}, and \ion{Fe}{2} lines are
labelled, as are the observed photospheric \ion{C}{2} and \ion{C}{3}
transitions. The symbol x indicates geocoronal emission lines, except
for the \ion{C}{3} line at 977 \AA\ that originates from scattered solar
emission.\label{fg:f1}}
\end{figure}

When searching for additional stellar lines, like \ion{C}{3} $\lambda
977$ and potential \ion{O}{1}, \ion{N}{1}, and \ion{N}{2} transitions,
it is necessary to extract the night-only data, because these lines can
be blended with strong emission lines coming from the terrestrial day
airglow (see \citet{feldmanetal01}). The night-only data are obtained
by reprocessing the raw data with the keyword DAYNIGHT set to NIGHT. In
this way, the FUSE pipeline processes only data that were taken during
the night. That subset amounts to 66 \% of the original
data secured.

\subsection{FOS and GHRS data}

Medium-resolution FOS red digicon data of \pga\ can, when combined with
medium-resolution GHRS data, be used to construct an energy
distribution which replaces the low signal-to-noise data secured many
years ago with the IUE satellite and repeatedly analyzed
since \citep{liebertetal86, tvs91, castanheiraetal06}.

The FOS data were acquired through the large ($3.7"\times 1.3"$)
aperture with the red G190H and G270H dispersers, and cover the region
between 1571 and 2311 \AA, and between 2222 and 3277 \AA\ at a
resolution $\lambda/\Delta\lambda=1300$. This corresponds to $\sim 1-2$
\AA\ through our wavelength range. These data were complemented with
medium-resolution GHRS data secured through two exposures with the
G140L grating and acquired at a spectral resolution of $\sim 0.65$
\AA\ (image number {\it z3eu0106p} in the 1136$-$1422 \AA\ range and
{\it z3eu0107p} in the 1389$-$1675 \AA\ range). At L$\alpha$, our GHRS
data are of lower quality than those secured at better spectral
resolution ($\sim 0.08$ \AA) by \citet {provencaletal00}. These data, a
log of which is given by \citet {provencaletal00}, were retrieved from
the MAST archives. The concatenated medium-resolution data are
displayed in Figure 2.

\begin{figure}[!ht]
\plotone{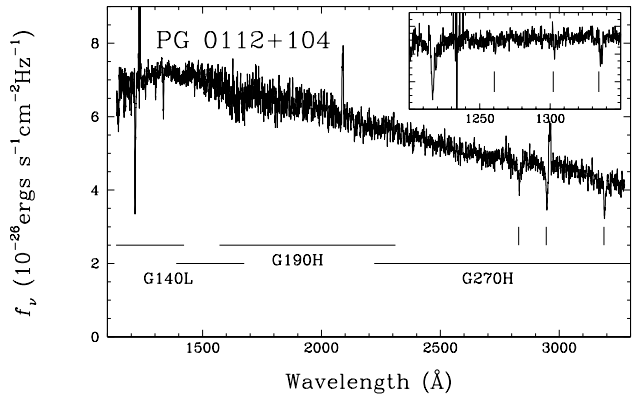}
\caption{Concatenated medium-resolution GHRS and
FOS data for \pga.  The data have been smoothed with a 10-point filter.
The ranges of the two exposures through the G140L grating on the GHRS
and of the two FOS exposures with the red G190H and G270H dispersers are
indicated. The three tick marks in the G270H range show the location of
strong \ion{He}{1} features, while the spikes near 2090\AA\ and 2960 \AA\
appear to be due to improper accounting for counter overflow. The box
features an unsmoothed blowup of the 1200$-$1350 \AA\ GHRS G140L region,
where the \ion{H}{1} L$\alpha$ line as well as the resonant interstellar
\ion{Si}{2} $\lambda$1260, \ion{O}{1} $\lambda$1302, and \ion{C}{2}
$\lambda$1334 lines are located. The locations of the last three are
indicated by tick marks. The four spikes in the 1230$-$1240 \AA\ range
appear to be associated with an on-board memory problem.\label{fg:f2}}
\end{figure}

\subsection{Optical spectrophotometry}

Two blue optical spectra of \pga\ were secured over a time span of 18
years (see Table~1 for a summary of our observations).  \pga\
initially belonged to a sample of over 100 DB stars observed over
several years in the blue optical region at the Steward Observatory
2.3~m Bok telescope. This sample formed the observational basis for
the work of \citet{phd}. The instrumental setup includes a Boller \&
Chivens Spectrograph, a 4.5 arcsec slit, and a 600 l mm$^{-1}$ grating
in f\/irst order. Together with a $800\times 800$ TI or a $1200\times
800$ Loral CCD, this combination provides coverage of the $\sim
3750-5100$ \AA\ region at an intermediate resolution of $\sim 6$
\AA. The \pga\ spectrum was characterized by a S/N ratio of
$120$. More recently, a new observation of \pga\ with the same
telescope was secured specifically for this paper.  The new data are
entirely consistent with the older spectrum secured in 1991, and have
the added feature that an annoying instrumental glitch present in the
older data near 3950 \AA\ has been removed.

\pga\ was also included in the sample of stars observed at H$\alpha$ and
analyzed by \citet {hunter01}. The red spectroscopy was secured during two
runs at the KPNO 4~m Mayall telescope equipped with the Ritchey-Chretien
Focus Spectrograph, UV Fast Camera, and T2KB CCD. The data from both
runs were later combined.  Coverage extends approximately from 5600 \AA\
to 7400 \AA, at a resolution of $\sim 3$ \AA. A S/N ratio of $\sim 90$
was achieved. The blue and red spectra are displayed together in Figure 3.

\begin{figure}[!ht]
\plotone{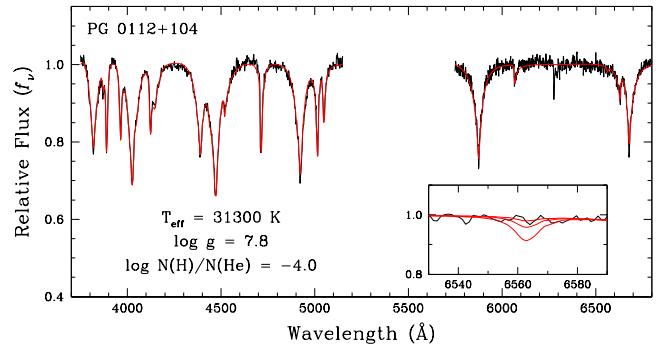}
\caption{Blue and red spectra for \pga. The optical
spectrum of that star shows only lines associated with the \ion{He}{1}
ion. The features near 6250 \AA\ are due to night-sky lines.
Superposed in red is our fit at the optimal parameters
discussed in the text. The insert shows the comparison of our predicted
H$\alpha$ profiles for hydrogen abundances of $\log N({\rm H})/N({\rm
He})=-4.0$, $-3.5$, and $-3.0$.\label{fg:f3}}
\end{figure}

\subsection{High Speed Photometry}

High speed photometry was obtained by GF and S.~Charpinet on 2002 July
14 UT with LAPOUNE, the portable Montr\'eal 3-channel photometer on the
CFHT 3.6~m ref\/lector. The instrument uses three Hamamatsu R647-04
photomultiplier tubes to measure simultaneously the target star, a
reference star, and the sky. The light curve of slightly over an hour
long (3650 s) was obtained under excellent photometric conditions, and
consists of 365 points sampled every 10~s. Our light curve is shown in
Figure 4.

\begin{figure}[!ht]
\plotone{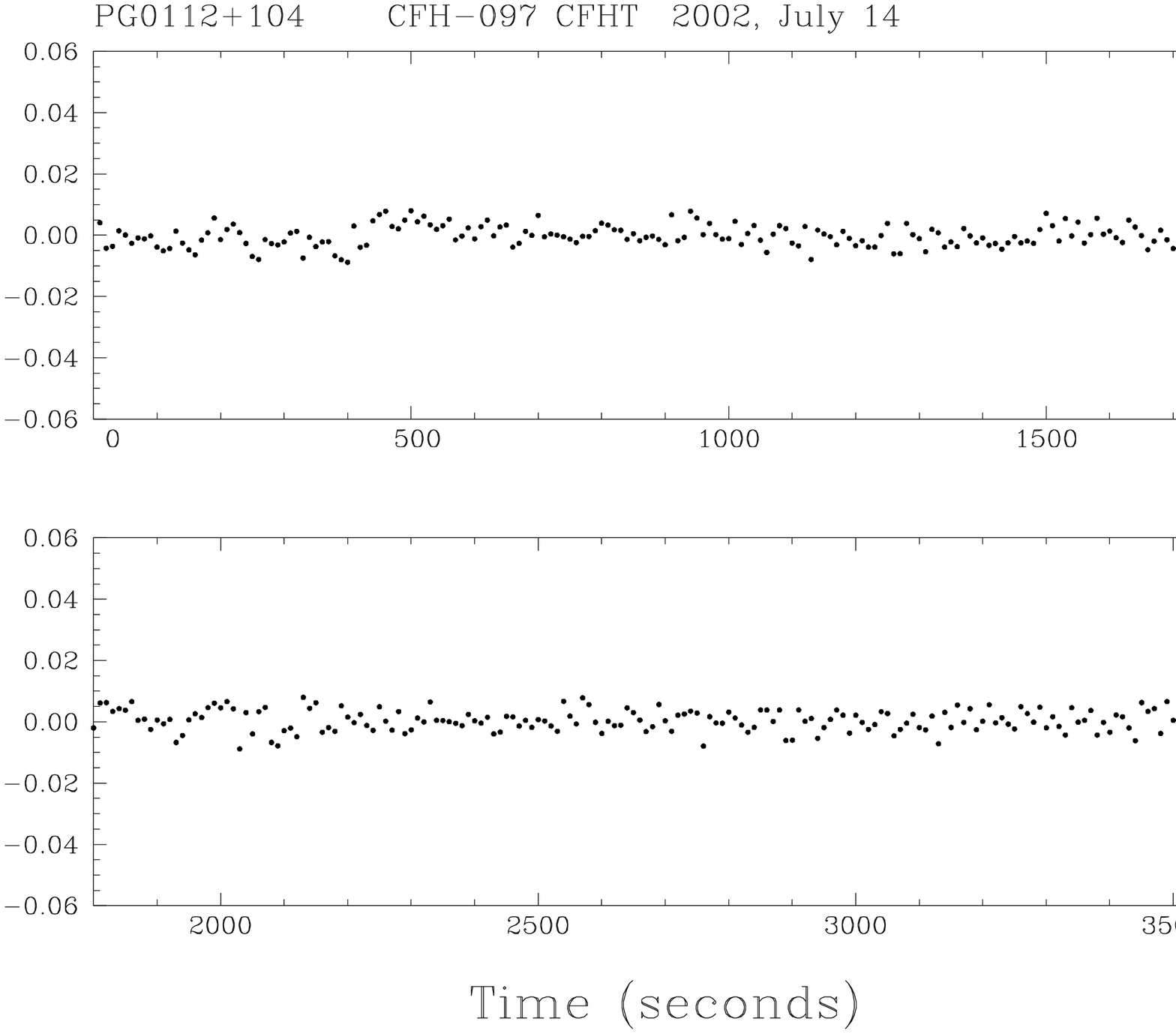}
\caption{Light curve of \pga\ obtained at the
CFHT. The observation lasted 3650~s, and the sampling time was
10~s.\label{fg:f4}}
\end{figure}

\section{Model atmosphere and synthetic spectrum calculations}

Our analysis is based on our updated grid of white dwarf models
described in \citet{tremblay09} in which we have incorporated the
improved Stark profiles of neutral helium of \citet{beauchamp97}.  In
the context of DB stars, these models are comparable to those described
by \citet{phd} and \citet{beauchamp96}, with the exception that at
low temperatures ($T<10,800$~K), we now use the free-free
absorption coefficient of the negative helium ion of \citet{john94},
but this should have no effect on the analysis of \pga\ (as discussed further
below). Our models are in LTE and include convective energy
transport within the mixing length theory. As in the analysis of
\citet{beauchampetal99}, we use here the parameterization described as
ML2/$\alpha=1.25$. Synthetic spectra based on these models were used
for the L$\alpha$ and optical spectra analyses.

In addition, several calculations were also carried out with TLUSTY and
SYNSPEC, the publicly-available model atmosphere codes developed by I.
Hubeny and one of us (TL) \citep[e.g.,][]{hl95}, which were also used
in the analysis of \citet{provencaletal00}. Specifically, we used these
codes for the determination of the carbon abundance and to set the
upper limits on other heavy elements (see \S\ 5.5 below) as well as to
reassess the importance of NLTE ef\/fects in DB stars (see \S\ 4
below). While the standard parameterization of the convective
ef\/ficiency within the mixing length theory included in TLUSTY
corresponds to that described by \citet{mihalas78}, we have modified
our version to be able to consider alternative efficiencies.

\section{The importance of NLTE Ef\/fects in DB Stars}

In their analysis of HST observations of helium-atmosphere white
dwarfs, \citet{provencaletal00} suggest that NLTE ef\/fects in the
continuum of DB stars might be significant and deserve consideration in
their analysis. Their Figure 3 shows, in particular, deviations at the
level of 9 \% near 1500 \AA, and at the level of 4 \% near 4000 \AA,
between emergent fluxes from a LTE model and those of a NLTE model,
both computed with TLUSTY at $T_{\rm eff}=25,000\,{\rm
 K}$. This comes a bit as a surprise since the early investigations of
both \citet{kudritzki76} and \citet{wesemael81} suggested that NLTE
ef\/fects in the continuum were negligible even at ef\/fective
temperatures considerably higher than that of \citet{provencaletal00}.
\citet{dw96} also justified a LTE analysis of the DB stars on the basis
of the complete absence of NLTE ef\/fects in the He I line spectrum at
$T_{\rm eff}=40,000\,{\rm K}$; NLTE ef\/fects in the lines can persist
even when continuum NLTE ef\/fects are completely negligible (a case in
point would be the sharp H$\alpha$ core in cool DA stars). Even more
recently, in a reanalysis of hot SDSS DB white dwarfs in the DB-gap,
\citet{HD09} concluded that LTE is a valid assumption for objects with
effective temperature below 45,000 K. The result of \citet{dw96} and
\citet{HD09} thus appear to confirm the earlier assessment that NLTE
ef\/fects are negligible for the analysis of DB stars.

Intrigued by the \citet{provencaletal00} claim, which is based on
TLUSTY models, we have computed our own NLTE models of DB stars with
TLUSTY. The input parameters are identical to those of
\citet{provencaletal00}, namely $T_{\rm eff}=25,000\,{\rm K}$, ${\rm
  log}\,g = 8.0$, ${\rm log}\,N({\rm C})/N({\rm He}) = {\rm
  log}\,N({\rm H})/N({\rm He}) = -5.0$, and the parameterization of
\citet{mihalas78} for the convective flux within the mixing-length
theory. We find no observable NLTE ef\/fects in the ultraviolet and
optical continua at that ef\/fective temperature. Furthermore,
additional tests suggest that no NLTE ef\/fects are apparent in the
optical He I line spectrum at 40,000 K and 30,000 K, the highest
temperature we considered. This is in complete agreement with the
result of \citet{dw96} and \citet{HD09}. However we find NLTE
ef\/fects within $\pm 2.5$ \AA\ of the core of the L$\alpha$ line,
which is formed high in the photosphere. We have explicitly checked
that these ef\/fects have no significant impact upon the analysis we
carry out, given that the saturated interstellar profile at ${\rm
  log}\,N_{\rm H} = 19.4$ dominates the core of the observed L$\alpha$
profile. We also explicitely verified that the models calculated with
small traces of carbon are identical to those without carbon. The
abundance of carbon needs to be several orders of magnitude higher,
which is ruled out by spectroscopic observations, to have any
detectable effect on the thermodynamic structure. Therefore, classical
LTE DB atmosphere models are sufficient for determining the effective
temperature and the surface gravity of \pga\ from the optical spectra.

\section{Analysis} 

\subsection{Determination of the atmospheric parameters from optical data}

The analyses of \citet{phd} and \citet{beauchampetal99} show that the
ef\/fective temperature determined from the optical spectrum of hot DB
stars depends on the hydrogen abundance adopted for the atmosphere,
despite the fact that hydrogen may not be directly visible. This is
why two ef\/fective temperatures, which dif\/fer by 200$-$4000 K, are
listed by \citet{beauchampetal99} for the majority of DB stars: the
hotter one was obtained under the assumption of a completely
hydrogen-free atmosphere, and the cooler one was obtained under the
assumption that hydrogen is present at an abundance slightly below
that at which the H$\beta$ line would become visible. In general,
thus, the fits to the He I lines in the optical provide a locus of
optimal ef\/fective temperatures as a function of assumed hydrogen
abundance.

The constraints that can be placed on the photospheric hydrogen
abundance thus represent important ingredients in the analysis of hot
DB stars. These can be obtained by searching either the H$\alpha$
transition in the red or the L$\alpha$ transition in the
ultraviolet. Both techniques have well-known advantages and
disadvantages. In rare cases, for example for bright or important
objects, both sets of data are available. This is the case for \pga.

The fits to the older blue optical spectrum of \pga\ by
\citet{beauchampetal99} yielded $T_{\rm eff} = 31,500$ K, $\log
g=7.82$ for a pure helium composition, and $T_{\rm eff} = 28,300$ K,
$\log g=7.76$ for a composition with $\log N({\rm H})/N({\rm He}) =
-3.0$. This representative abundance was chosen at the time on the
basis of a quick look at the newly-acquired GHRS data which appeared
to show a fairly strong L$\alpha$ profile. Of course, the profile has
since been shown by \citet{provencaletal00} to have a substantial
interstellar contribution, and their derived photospheric hydrogen
abundance, namely $-4.0 < \log N({\rm H})/N({\rm He}) < -3.5$, is
consequently somewhat lower than the nominal value used by
\citet{beauchampetal99}.

As an internal check of our updated model grid discussed in \S~3, we
analyzed this older spectrum using our fitting technique that relies
on the nonlinear least-squares method of Levenberg-Marquardt
\citep{press86}, which is based on a steepest descent method. The
model spectra (convolved with a Gaussian instrumental profile) and the
optical spectrum are first normalized to a continuum set to unity;
this continuum is set in a similar fashion to that described in detail
in \citet{liebert05} for DA stars. The calculation of $\chi ^2$ is
then carried out in terms of these normalized line profiles
only. Atmospheric parameters -- $T_{\rm eff}$, $\log g$ -- are
considered free parameters in the fitting procedure; in the case of
\pga\ we assume a given value of the hydrogen abundance and compare
the predicted H$\alpha$ profile with the observations to set a limit
on the hydrogen abundance (see below). Our atmospheric parameters with
this old spectrum yields $T_{\rm eff} = 31,600$~K and $\log g=7.85$
under the assumption of a pure helium composition, in excellent
agreement with the values reported by \citet{beauchampetal99}, which
suggests that both model grids for DB stars are entirely consistent.

Since the analysis of \citet{beauchampetal99}, a new blue spectrum as
well as spectra covering H$\alpha$ have been secured. Our analysis of
this blue spectrum using our pure helium grid yields $T_{\rm eff} =
31,590$~K and $\log g=7.82$, in almost perfect agreement with the
results obtained with our old spectrum. Our procedure is then to fit
this spectrum assuming various hydrogen abundances and compare the
predictions at H$\alpha$ with our red spectrum. These comparisons are
shown in the insert of Figure \ref{fg:f3} for $\log N({\rm H})/N({\rm
  He})=-4.0$ (31,320~K, 7.82; this is almost identical to our pure
helium solution), $-3.5$ (30,780~K, 7.81), and $-3.0$ (29,140~K,
7.78). Since H$\alpha$ is clearly not detected in our red spectrum, we
adopt as our final solution $T_{\rm eff} = 31,300\pm 500$~K, $\log g =
7.8\pm 0.1$, and ${\rm log}\,N({\rm H})/N({\rm He})\leq -4.0$, a
hydrogen abundance limit that not is grossly inconsistent with the
result of \citet{provencaletal00} based exclusively on the L$\alpha$
profile.  Our adopted solution is superposed in Figure \ref{fg:f3} on
top of the the blue and red spectra.

\subsection{Constraint on the hydrogen abundance from the L$\alpha$ profile}

When L$\alpha$ observations are available, the contribution of both the
geocoronal feature and the interstellar medium absorption must be
allowed for. As shown by \citet{provencaletal96} and
\citet{provencaletal00} in their pioneering analyses of the L$\alpha$
profile in \pga, GD 190, and in the prototypical V777 Her star GD 358, this can
be done successfully when the saturated core and the broad wings
are considered separately. For nearby objects, this allows limits on
the photospheric hydrogen content to be placed.

The values of the atmospheric parameters we have secured on the basis
of the optical data in the preceding section are likely to be the most
reliable in terms of the assumptions underlying our analysis.
Nevertheless, other spectral ranges, perhaps less amenable to detailed
analyses, can contribute to the overall self-consistency of the
analysis. In this spirit, we have investigated the consistency of the
L$\alpha$ profile, following the method and technique outline by
\citet{provencaletal00} in their original analysis, with our optical
determination of the atmospheric parameters of \pga. The saturated
part of the profile suggests a fairly well constrained interstellar
hydrogen column density of ${\rm log}\,N_{\rm H} = 19.4\pm 0.1$, a
value consistent with the results of the initial analysis. The
resulting photospheric hydrogen abundance depends on the assumed
ef\/fective temperature, and --- given the importance assumed by the
interstellar core --- a two-dimensional fit would likely not be
significant. We plot, in Figure 5, the match to the overall
ISM$+$photospheric L$\alpha$ profile secured with models at our
optically determined effective temperature and gravity. This is done
for values of the hydrogen abundance of ${\rm log}\,N({\rm H})/N({\rm
  He})=-3.0$, $-4.0$, and $-5.0$. The two panels display matches
secured with values bracketing the optimal column density determined
above.  The theoretical profiles become less sensitive to the hydrogen
abundance for ${\rm log}\,N({\rm H})/N({\rm He}) < -4.0$, and an
accurate determination of the hydrogen content from the L$\alpha$
profile seems hardly possible. Suffice it to be said that the value we
derive from the optical spectrum is consistent with the limit set from
the HST data.

\begin{figure}[!ht]
\plotone{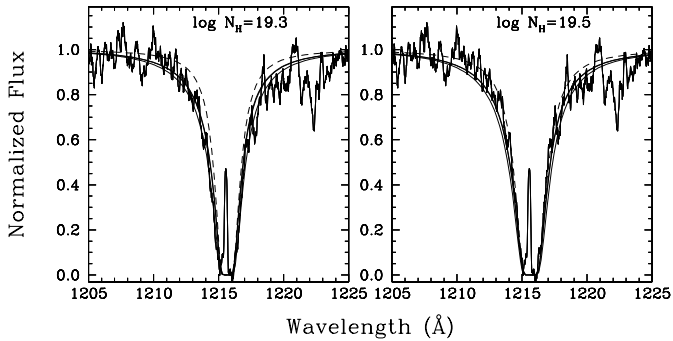}
\caption{Match to the L$\alpha$ spectrum of \pga\
for two labeled column densities, that bracket our best-fitting value,
${\rm log}\,N_{\rm H} = 19.4\pm 0.1$. The effective temperature is $T_{\rm
eff} = 31,300$ K, the surface gravity is log $g = 7.8$, and the hydrogen
abundances are ${\rm log}\,N({\rm H})/N({\rm He})$ from $-3.0$ to $-5.0$
in steps of $1.0$, from the outside to the inside. The models at $-4.0$
and $-5.0$ are nearly on top of each other. In both panels, the innermost
short dash profile is the purely interstellar profile.\label{fg:f5}}
\end{figure}

\subsection{Transitions of \ion{He}{1} and \ion{He}{2}}

Our FOS data show several members of the \ion{He}{1} far-ultraviolet
series originating on the $2^3{\rm S}$ lower level at $E = 159850\,{\rm
cm}^{-1}$. Prominent among those are the first members at 3187.74
\AA\ ($2^3{\rm S}-4^3{\rm P}$), 2945.10 \AA\ ($2^3{\rm S}-5^3{\rm P}$),
and 2829.07 \AA\ ($2^3{\rm S}-6^3{\rm P}$). These transitions can been
seen faintly in some of the better exposed IUE LWR spectra of DB white
dwarfs \citep{holbergetal03}. As was done with the L$\alpha$
profile, we can investigate here the rough consistency of our
atmospheric parameters with those new features, amenable to a
preliminary analysis. In Figure 6, we plot the observed FOS data
together with a synthetic spectrum generated with SYNSPEC at the values
of the atmospheric parameters determined above. While the match to the
first two transitions appears quite good, it worsens as we move up the
series. This is undoubtedly due to a combination of factors, namely
the lack of tabulated electron impact widths for the higher line
members and the omission, within SYNSPEC, of detailed treatment of the
level dissolution as we approach the series limit near 2644 \AA.

\begin{figure}[!ht]
\plotone{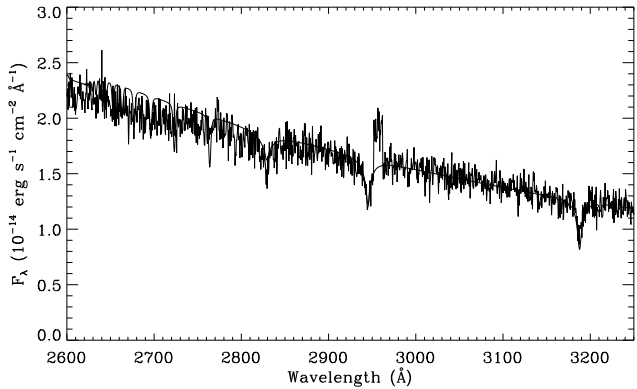}
\caption{Portion of the FOS spectrum showing the
\ion{He}{1} transitions at 2829.07, 2945.10, and 3187.74 \AA\ that
originate on the $2^3{\rm S}$ lower level at $E = 159850\,{\rm
cm}^{-1}$, together with a synthetic spectrum generated by
SYNSPEC.\label{fg:f6}}
\end{figure}

The S/N ratio of the medium-resolution GHRS data around the \ion{He}{2}
$\lambda 1640$ transition is unfortunately too low to provide a
useful check on the atmospheric parameters of \pga. Our synthetic
spectra do predict a weak \ion{He}{2} $\lambda 1640$ transition at this
effective temperature, but no meaningful confrontation with the data is
possible.

\subsection{Energy distribution}

The spectral energy distribution of the star provides an additional
piece of information. While its slope is, for classical hot DB stars,
relatively insensitive to the photospheric hydrogen content and to the
surface gravity, it displays some sensitivity to the ef\/fective
temperature, especially when the energy distribution can be sampled as
far as possible in the ultraviolet. While issues related to
interstellar reddening and calibration must then be addressed, the
matching of the energy distribution of \pga\ affords us with another
consistency check on the adopted values of $T_{\rm eff}$ and ${\rm
log}\,N({\rm H})/N({\rm He})$ which cannot be overlooked. Figure 7
displays the energy distribution we have constructed for \pga. It
includes the FUSE data, sampled in 20\AA-wide bins, the HST and GHRS
data, sampled in 40\AA-wide bins, as well as the multichannel data,
converted to monochromatic flux according to the relation of
\citet{jlg76}. The match at $V(1.85\,\mu{\rm m}^{-1}$) defines the
solid angle for all the data.

\begin{figure}[!ht]
\plotone{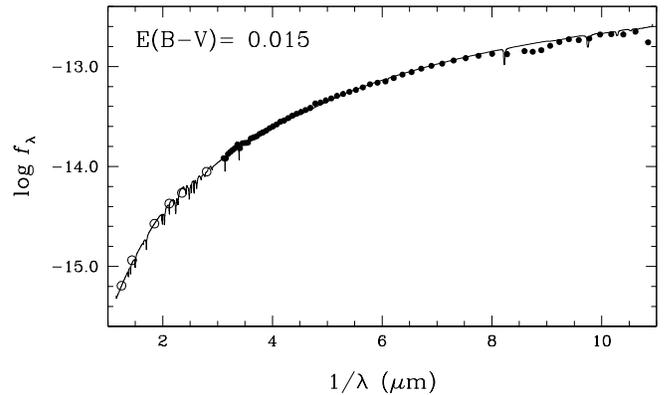}
\caption{The spectral energy distribution of our
optimal model (solid line), together with the binned FUSE, GHRS, FOS
spectra (filled circles) and the six multichannel magnitudes from
\citet{jlg84} (open circles). All data are normalized to the $V$
magnitude. The model parameters are $T_{\rm eff} = 31,300$ K, log $g =
7.8$, ${\rm log}\,N({\rm H})/N({\rm He}) = -4.0$. Interstellar reddening is
included, with $E(B-V)=0.015$.\label{fg:f7}}
\end{figure}

There are, for this object, two modern sets of magnitudes available.
The first one, by \citet{jlg84}, includes the multichannel $V$
magnitude ($V=15.36$) together with five MCSP color indices from which six
monochromatic magnitudes can be extracted. The second data set,
provided by \citet{jlgjl90}, lists a full set of 15 monochromatic
magnitudes, all simply read off CCD spectrophotometric data. The visual
magnitude obtained in that last effort, $V=15.51$, is a full 0.15 mag
fainter than that provided by \citet{jlg84}. On the basis of our
experience with these data, we restrict our analysis here to the full
MCSP data of 1984, which is given on the AB79 calibration scale of
\citet{og83}. Our procedure includes a provision for the contribution
of interstellar reddening, as described by the analytic expression for
interstellar extinction of \citet{seaton79}.

Overall, the \citet{jlg84} data leads to a reasonable fit to the energy
distribution in the long-wavelength ultraviolet range. The FUSE data,
on the other hand, are below the predicted model fluxes. An improved
match can be secured by including modest amounts of reddening $E(B-V)
\sim 0.03$. This value is consistent with the low column density
of neutral hydrogen determined earlier on the basis of the L$\alpha$
profile, namely ${\rm log}\,N_{\rm H} = 19.4\pm 0.1$, as well as with
the apparent absence of ${\rm H}_2$ absorption in the FUSE spectra.

\begin{deluxetable*}{lcrcrccc}
\tablecolumns{8}
\tablewidth{0pc}
\tablecaption{Carbon Line Equivalent Widths and Abundances Observed in FUSE and GHRS Spectra}
\tablehead{
\colhead{Ion} & \colhead{$\lambda$ (\AA)}
& \colhead{log $gf$} & \colhead{$g_{\ell}$}
& \colhead{$E_{\ell} ({\rm cm}^{-1})$} & \colhead{E.W. (m\AA)}
& \colhead{Abundance} 
& \colhead{Instrument} \\
}
\startdata
\ion{C}{3}   & $977.020$ & $-0.120$ & $1.0$ &     $0.000$  &  $165.8\pm 13.2$  & $-5.85\pm 0.13$     & FUSE \\
&&&&&& \\
\ion{C}{2}   & $1009.858$ & $-0.457$ & $2.0$ &  $43003.300$  &  $135.3\pm 12.2$  & $-6.13\pm 0.08$  & FUSE \\
      & $1010.083$ & $-0.156$ & $4.0$ &  $43025.300$ \\
      & $1010.371$ & $ 0.200$ & $6.0$ &  $43053.600$ \\
&&&&&& \\
\ion{C}{2}   & $1036.337$ & $-0.611$ & $2.0$ &     $0.000$  &  $63.7\pm 4.7$   & $-6.03\pm 0.12$    & FUSE \\
      & $1037.018$ & $-0.310$ & $4.0$ &    $63.420$  & $82.1\pm 6.2$ \\
&&&&&& \\
\ion{C}{2}   & $1065.891$ & $ 0.001$ & $6.0$ & $74930.100$  & $23.5\pm 4.9$   & $-6.22\pm 0.14$    & FUSE \\
      & $1065.920$ & $-0.952$  & $4.0$ & $74932.620$ \\
      & $1066.133$ & $-0.255$  & $4.0$ & $74932.620$ \\
&&&&&& \\
\ion{C}{3}  & $1174.933$ & $-0.468$  & $3.0$ & $52390.750$  & $289.4\pm 19.0$   & $-6.54\pm 0.10$   & FUSE \\
      & $1175.263$ & $-0.565$  & $1.0$ & $52367.060$ \\
      & $1175.590$ & $-0.690$  & $3.0$ & $52390.750$ \\
      & $1175.711$ & $ 0.009$  & $5.0$ & $52447.110$ \\
      & $1175.987$ & $-0.565$  & $3.0$ & $52390.750$ \\
      & $1176.370$ & $-0.468$  & $5.0$ & $52447.110$ \\
&&&&&& \\
\ion{C}{2}   & $1323.862$ & $-1.296$  & $6.0$ & $74930.100$ &  .........          && GHRS \\
      & $1323.906$ & $-0.342$  & $4.0$ & $74932.620$ &  ......... \\
      & $1323.951$ & $-0.150$  & $6.0$ & $74930.100$ &  ......... \\
      & $1323.995$ & $-1.297$  & $4.0$ & $74932.620$ &  ......... \\
&&&&&& \\
\ion{C}{2}   & $1334.532$ & $-0.597$  & $2.0$ &     $0.000$ &   $81.4\pm 8.1$   & $-6.14\pm 0.15$    & GHRS \\
      & $1335.663$ & $-1.295$  & $4.0$ &    $63.420$ &  $119.7\pm 9.8$ \\
      & $1335.708$ & $-0.341$  & $4.0$ &    $63.420$ \\
\enddata
\end{deluxetable*}

The parallax of \pga, $\pi = 8.77 \pm 0.55\,{\rm mas}$ \citep{hh09},
provides an additional consistency check on our analysis. Our preferred
match based on the $V$ magnitude of \citet{jlg84} and an interstellar
reddening of $E(B-V)= 0.015$ yields a solid angle of $\Omega \equiv \pi
R^2/D^2 = 2.406\times 10^{-23}$ and a radius of $R=0.0140\,{R}_\odot$.
Here, we have assumed $T_{\rm eff} = 31,300$ K and a composition with
${\rm log}\,N({\rm H})/N({\rm He}) = -4.0$. We now use the \citet{w92}
models with a carbon core, a helium envelope of $10^{-4}\,M_\star$,
and no hydrogen layer, representative of helium-atmosphere white
dwarfs. On the basis of these models, we derive log $g = 7.90\pm
0.16$ for our energy distribution match, a value consistent with the
spectroscopically-determined value.

\subsection{Abundances of heavy elements}

In addition to interstellar features associated with \ion{C}{2},
\ion{C}{3}, \ion{N}{1}, \ion{N}{2}, \ion{O}{1}, \ion{Si}{2}, and
\ion{Fe}{2}, the FUSE spectra of \pga\ exhibit many of the photospheric
carbon features previously seen in other hot DB stars by
\citet{petitclercetal05} and \citet{desharnaisetal08}. Five transitions
were observed: the \ion{C}{2} $\lambda$1010 triplet, the \ion{C}{2}
$\lambda$1036 and $\lambda$1066 doublets, and the \ion{C}{3}
$\lambda$977 line and $\lambda$1175 complex. Among those, the two
components of the \ion{C}{2} doublet, $\lambda$1036.337 and
$\lambda$1037.018, are split since the doublet originates on levels of
low excitation energy (ground-state for the blue component, and 63.4
cm$^{-1}$ above the ground state for the red component). Both doublet
components thus exhibit well-separated contributions originating in the
photosphere and in the ISM. For the other carbon features observed, the
energy of the lower levels is high enough to guarantee a photospheric
origin. Table 2 lists the transitions uncovered and equivalent widths
measured in the FUSE spectra of \pga. Among the DB stars previously
studied with FUSE, EC 20058$-$5234 --- with its ef\/fective temperature
near 28,000 K --- is the closest analog to \pga: in that object,
\citet{petitclercetal05} reported the \ion{C}{2} $\lambda$1010 and
$\lambda$1036 doublets, as well as the \ion{C}{3} $\lambda$1175
complex. The \ion{C}{2} $\lambda$1066 doublet was not observed. In
addition, the reanalysis by \citet{dwb02} of the GHRS spectra of
\pga\ secured by \citet{provencaletal00} had shown the presence of the
\ion{C}{2} $\lambda$1335 doublet in those data.

We have redetermined individual carbon abundances on the basis of the
six transitions of \ion{C}{2} and \ion{C}{3} observed in the ultraviolet
spectra of PG~0112+104. The results are summarized in Table~2 and shown
in Figure~8. Our analysis differs little from those carried out earlier
(Dufour et al. 2002; Petitclerc et al. 2005; Desharnais et al. 2008). We
abstained from carrying out a detailed analysis of the interstellar
components in the \ion{C}{2} doublets. In the case of the \ion{C}{3}
$\lambda$977 line, we analyzed the night-time data of the SiC channels
in order to avoid the scattered solar emission contribution. For each
transition, the individual uncertainty on the carbon abundance ranges
from $\pm0.08$ dex to $\pm0.15$ dex. The uncertainties on the carbon
abundances take into account uncertainties on the quality of the fit,
uncertainties on the oscillator strengths, and uncertainties on the
atmospheric parameters ($\Delta T_{\rm{eff}} = 500$~K, $\Delta \log g =
0.1$ dex, and $\Delta {\rm{H/He}} = 0.3$ dex). Table 2 shows that the
carbon abundances vary from $\log N({\rm{C}})/N({\rm{He}}) = -5.85$ to
$-6.54$. These two extremes come from the two \ion{C}{3} $\lambda$977
and $\lambda\lambda$1175 lines. Considering all carbon abundances,
we obtain a mean carbon abundance of $\log N({\rm{C}})/N({\rm{He}}) =
-6.15\pm0.23$, where the uncertainty is the standard deviation of the
measurements. This abundance is consistent with that obtained by Provencal
et al. (2000) on the basis of the \ion{C}{2} doublet in their GHRS data,
$\log N({\rm{C}})/N({\rm{He}}) = -5.8\pm0.3$ (determined for a slightly
cooler effective temperature near 30,000~K), and with the value estimated
by Dufour et al. (2002) from the same data, $\log N({\rm{C}})/N({\rm{He}})
= -6.0$ (for the same effective temperature near 30,000~K).

\begin{figure}[!ht]
\plotone{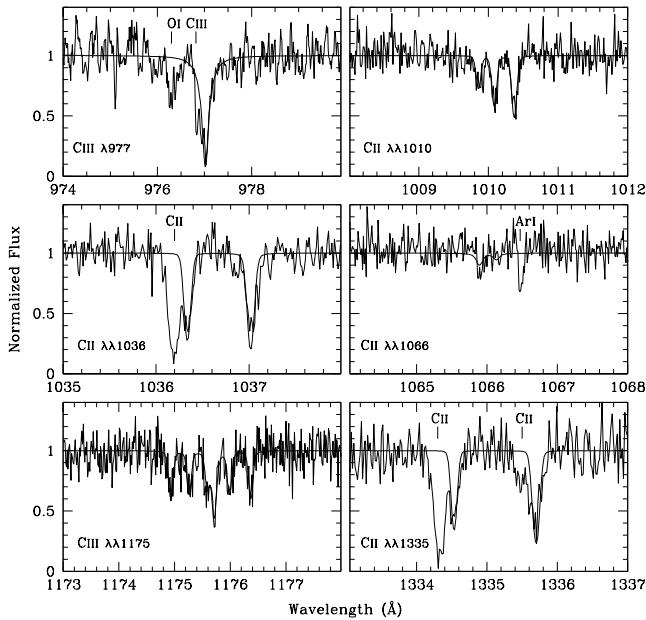}
\caption{Fits to the carbon transitions observed in
the FUSE and GHRS range. The \ion{C}{2} $\lambda$1036 and $\lambda$1335
features, which originate on low-lying levels, both exhibit a contribution
formed in the ISM. Two interstellar \ion{O}{1} and \ion{Ar}{1} lines
are also labelled.  \label{fg:f8}}
\end{figure}

We inspected the {\it FUSE} spectrum in detail to search for the presence
of other elements in the atmosphere of PG~0112+104. No photospheric lines
other than carbon were detected.  Given that the {\it FUSE} wavelength
range is rich in resonance lines or strong transitions, we were able to
estimate abundance upper limits of a dozen elements.  Table 3 summarizes
our results and shows the lines that we used for determining the upper
limits. The upper limits correspond to abundances that are 3$\sigma$
above the lowest detectable abundances. We measured upper limits of
six light elements (N, Si P, S, Cl, and Ar), five iron peak elements
(V, Cr, Mn, Fe, and Co), and one element beyond the iron peak (Pb). The
upper limits range from $-7.0$ to $-8.7$.  Elements such as oxygen and
calcium do not have strong lines in the {\it FUSE} pass band for a star
like PG~0112+104, so no upper limits could be set for these elements.

The problem posed by the presence of carbon in hot, classical DB
stars such as GD 358, EC 20058$-$5234 or \pga\ was addressed by
\citet{desharnaisetal08}, who used the carbon abundance value published
by \citet{provencaletal00} (${\rm log}\,N({\rm C})/N({\rm He}) =
-5.8\pm 0.3$) in their discussion. Our updating of the carbon abundance
in \pga\ does little to alleviate the problems discussed there and in
\citet{petitclercetal05}: ``hot'' DB stars are too cool for significant
radiative element support, but too hot for significant dredge-up or
accretion from the ISM.

\citet{fontainebrassard05} have proposed an attractive scenario to
account for the presence of carbon in DB stars above 23,000 K.  In
their view, the observed carbon abundances result from the competition
in hot DB stars of efficient downward gravitational settling with a
weak stellar wind left over from previous evolution.  While there is
currently no direct evidence for this wind, its existence is not
unreasonable given that signs of mass loss are observed in the PG 1159
progenitors of hot DB stars. The investigation of winds along the
cooling track performed by \citet{fontainebrassard05} is based on
fully evolutionary models which include a linear relationship between
mass loss and age, with the wind turning off completely by the time a
star reaches $T_{\rm eff} = 20,000$ K. For the hot DB stars, the
resulting carbon abundance depends, in the main, on the mass loss rate
and is not strongly dependent on the structural parameters of the star
(for example, on the helium layer thickness). To account for the
abundances summarized by \citet{desharnaisetal08}, the rates invoked
are of the order of a few $10^{-13}$ M$_{\rm \odot}$ yr$^{-1}$.

Large amounts of carbon are also found in the so-called hot DQ white
dwarfs \citep{dufour07,dufour08}. These carbon-dominated atmosphere white
dwarfs are believed to be the offsprings of DB white dwarfs with very thin
helium layers that have experienced a convective mixing episode with
the underlying carbon envelope. However, the hot DQ stars are all found at
a much lower effective temperatures ($\sim 18,000-24,000$~K) than \pga.
It thus appears very unlikely that the carbon observed in the atmosphere
of this hot DB white dwarf could be the result of a similar convective
transformation caught in its early phase.

\begin{deluxetable}{lrrcrc}
\tablecolumns{6}
\tablewidth{0pc}
\tablecaption{Upper Limits on the Abundances of Heavy Elements
\label{tab:upper_limits}}

\tablehead{
\colhead{Ion} &
\colhead{$\lambda$ (\AA)} &
\colhead{$\log gf$} &
\colhead{$g_l$} &
\colhead{$E_l$ (cm$^{-1}$)} &
\colhead{$\log N({\rm{X}})/N({\rm{He}})$} 
}
\startdata
\ion{N}{3}  &  991.511 & $-1.317$ & 4 &   174.400 & $<-7.0$ \\
            &  991.577 & $-0.357$ & 4 &   174.400 &         \\
\ion{Si}{3} & 1113.174 & $-1.356$ & 5 & 53115.012 & $<-8.7$ \\
            & 1113.204 & $-0.186$ & 5 & 53115.012 &         \\
            & 1113.230 &  $0.564$ & 5 & 53115.012 &         \\
\ion{P}{3}  & 1003.600 & $-0.400$ & 4 &   559.140 & $<-7.8$ \\ 
\ion{S}{4}  & 1072.996 & $-0.829$ & 4 &   951.100 & $<-7.2$ \\
\ion{Cl}{3} & 1015.022 &  $0.050$ & 4 &     0.000 & $<-8.3$ \\
\ion{Ar}{2} &  932.054 &  $0.120$ & 4 &  1431.580 & $<-7.4$ \\
\ion{V}{3}  & 1149.945 &  $0.068$ & 10&   583.800 & $<-7.6$ \\
\ion{Cr}{3} & 1033.232 & $-0.197$ & 9 &   576.080 & $<-7.2$ \\
            & 1033.433 & $-0.259$ & 7 &   356.550 &         \\
            & 1033.680 & $-0.245$ & 7 &   356.550 &         \\
\ion{Mn}{3} & 1108.164 & $-0.057$ &12 & 26824.400 & $<-7.4$ \\
            & 1111.104 & $-0.169$ &12 & 26851.100 &         \\
            & 1113.186 & $-0.293$ & 8 & 26859.900 &         \\
\ion{Fe}{3} & 1122.526 & $-0.149$ & 9 &     0.000 & $<-7.4$ \\
\ion{Co}{3} &  939.062 & $-0.047$ &10 &     0.000 & $<-7.2$ \\
\ion{Pb}{3} & 1048.877 &  $0.114$ & 1 &     0.000 & $<-8.7$ \\
\enddata
\end{deluxetable}

\subsection{Is \pga\ a True V777 Her Variable?}

\pga\ had generally been considered a constant star on the basis
high-speed photometry carried out by \citet{rw83} and \citet
{kawaleretal94}. Its ef\/fective temperature has generally been
located in the 27,000 K -- 30,000 K range \citep{liebertetal86,
  beauchampetal99, provencaletal00, castanheiraetal06}, although
\citet{tvs91} quote a temperature as low as 24,300 K. 

Because its ef\/fective temperature placed it early on near the blue
edge of the instability strip of the V777 Her variables, high-speed
photometry of \pga\ has been carried out on several occasions to
search for the non-radial pulsations that characterize the V777 Her
stars.  Only upper limits on variations were reported. Thus
\citet{rw83} list maximum semi-amplitudes of 0.30 \% in the 10~s$-$
50~s window, of 0.21 \% in the 50~s$-$ 200~s window, and of 0.29 \% in
the 200~s$-$ 1200~s window, while \citet {kawaleretal94} report upper
limits on the amplitude of brightness variations of 0.34 \%, the later
being based on UV data where the amplitudes are expected to be even
larger than in the optical.

More recently, however, \citet{shipmanetal02} and \citet{jlp03} have
reopened the debate and suggested, on the basis of 30 h of observing
carried out in 2001 October at the McDonald observatory 2.1-m
telescope, that the object might be variable, with two periods
tentatively identif\/ied: $P = 168.98\,{\rm s}$, with an amplitude of
0.083 \%; and $P = 197.76\,{\rm s}$, with an amplitude of 0.087
\%. The noise level in the Fourier amplitude spectrum is $\sim$ 0.018
\%. The periods they infer are consistent with those that characterize
known V777 Her stars ($100-1100$ s), but the reported amplitudes are
considerably lower than those observed up to now even in the
lowest-amplitude V777 Her pulsators, which are $\sim$ 4.6 \% (for PG
1351$+$489 and PG 2246$+$121).

If the pulsations observed by \citet{shipmanetal02} in \pga\ are real,
it is unclear to what extent the driving mechanism --- and its
associated instability strip --- invoked for the regular V777 Her
stars may be relevant to the variations they observed. While the
\citet{shipmanetal02} and \citet{jlp03} data were considered
preliminary, their potentially crucial importance within the context
of pulsation theory prompted us to observe \pga\ as a backup object
during one of our high-speed photometry run at the CFHT.

As shown in Figure 9, white light brightness variations with amplitudes
greater than 0.08 \% of the mean brightness of the star can be ruled
out in the period window from 20 s to 1800 s, a range that includes all
known periodicities in DBV variables. With a noise level of the order
of twice that of \citet{shipmanetal02}, we see no structure at the
frequency (5.06~mHz) corresponding to the 198~s period, while a noise
peak might be present at the frequency (5.92~mHz) corresponding to the
169~s period detected by \citet{shipmanetal02}. Our upper limits on
the brightness variations in \pga\ are comparable to (even slightly
lower than) the variations reported by \citet{shipmanetal02}, so we
cannot confirm their suggested detection. We note in this context that
had \pga\ been a normal V777 Her variable, our observations would have
revealed it easily.

\begin{figure}[!ht]
\plotone{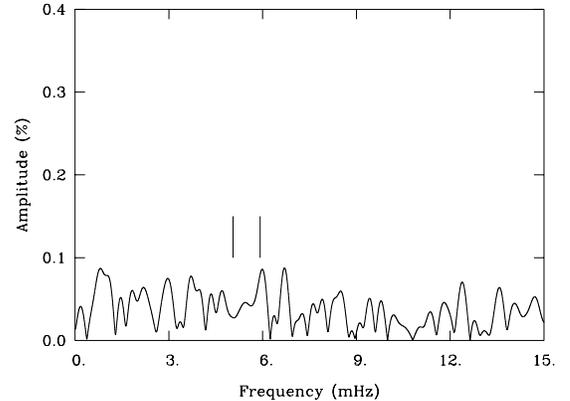}
\caption{Fourier transform of the light curve of
\pga.  No peak higher than 0.08 \% of the mean brightness is observed
in the period window from 20 s to 1800 s.\label{fg:f9}}
\end{figure}

What is, then, the status of the variations observed by
\citet{shipmanetal02}? To our knowledge, neither was their preliminary
report published in a refereed journal nor was their claim withdrawn.
It is assuredly of some significance that \pga\ was not considered
a V777 Her star in the recent review of \citet{wk08}. Even more
recently, the discussion of the instability strip of the DB stars of
\citet{nittaetal09} includes a private communication from J. Provencal
stating that ``time series observations of this star [\pga] have not
detected any pulsations''. This revised position is in agreement with
our own result, and suggests that \pga\ truly sits above the blue edge
of the V777 Her instability strip.

\section{Conclusions}

We have provided a comprehensive multiwavelength analysis of the hot
DB white dwarf \pga\ based on data largely unexploited up to now. Our
analysis yields the following parameters: $T_{\rm eff} = 31,300\pm
500$ K, log $g = 7.8\pm 0.1$ ($M=0.52\,M_\odot$), a hydrogen abundance
of ${\rm log}\,N({\rm H})/N({\rm He}) < -4.0$, and an improved carbon
abundance of ${\rm log}\,N({\rm C})/N({\rm He}) = -6.15\pm
0.23$. Since the discovery of several hot DB stars within the SDSS has
contributed to the demise of the idea of a true DB gap along the white
dwarf cooling sequence, the individual modeling of a single ``hot'' DB
white dwarf may have lost some of its luster. Nevertheless, at $V\sim
15.4$, \pga\ remains one of the few hot DB stars currently amenable to
a detailed analysis of the type completed here. \pga\ represents our
best object near the transition from nonpulsator to pulsator and our
reevaluation of the status of \pga\ as a V777 Her star in \S\ 5.6
underscores the fact that, even for the brightest members, a complete
picture of these complex objects is only slowly developing. There are
still many aspects about the evolution of helium atmosphere white
dwarf that are not completely understood yet, for instance the
observed carbon and hydrogen abundance patterns and the exact location
of the instability strip. This study brings much needed empirical
knowledge about these issues but more work on helium-rich white dwarfs
will be needed for a complete understanding of these objects to emerges.

\section{Acknowledgements}

We are grateful to S. Charpinet, R. Lamontagne, D. Paquin-Ricard, and
G. Scarpa for their contributions at various stages of this project,
to D. Durand and M.R. Rosa for cogent comments on the HST data, to
J.L. Provencal for sharing her spectroscopic and model atmosphere results
with us along the way, and to H. Harris and to the U.S. Naval Observatory
for making the parallax available in advance of publication. PD is a
CRAQ postdoctoral fellow and PC was a Canadian representative to the
FUSE project supported by CSA under a PWGSC contract. TL was supported
by NASA grant NAG5-12480, and PB is a Cottrell Scholar of the Research
Corporation for Science Advancement. This work was supported in part by
the NSERC Canada and by the Fund FQRNT (Qu\'ebec).

\end{document}